\documentclass{emulateapj}
\usepackage{amsmath,amssymb}

\shorttitle{Antiprotons Produced in SNRs}
\shortauthors{Berezhko \& Ksenofontov}

\begin{document}
\title{Antiprotons Produced in Supernova Remnants}

\author{E. G. Berezhko and L. T. Ksenofontov}
\altaffiltext{}{Yu. G. Shafer Institute of Cosmophysical Research and 
Aeronomy, 31 Lenin Ave., 677891 Yakutsk, Russia}

\email{ksenofon@ikfia.sbras.ru}

\begin{abstract}
We present the energy spectrum of antiproton cosmic ray (CR)
component calculated on the basis of the
nonlinear kinetic model of CR production in supernova remnants (SNR).
The model includes reacceleration of already existing in 
interstellar medium  antiprotons 
as well as  creation of antiprotons in nuclear
collisions of accelerated protons with gas nuclei 
and their subsequent acceleration by SNR shock.
It is shown that antiprotons production in
SNRs produces considerable effect in their resultant energy spectrum 
 making it essentially flatter above 10~GeV
so that the spectrum at TeV-energies increases by a factor of five.
Calculated antiproton spectrum is well consistent with the
PAMELA data, which correspond to energies below 100~GeV. 
As a consistency check  
we have also calculated
within the same model the energy spectra of secondary nuclei
and show that the measured  boron-to-carbon ratio 
is consistent with the significant SNR contribution.
\end{abstract}

\keywords{acceleration of particles --- cosmic rays --- ISM: supernova remnants}

\section{Introduction}

There is a great astrophysical interest in cosmic ray (CR) antiprotons.
It is believed that most of antiprotons originate in collisions
of CR protons with interstellar medium (ISM) gas nuclei.
Therefore antiprotons represent a kind of 
so-called secondary CR component, opposite to the primary CRs, which
originate in CR sources, presumably in supernova remnants (SNRs).
The same is trough for positrons, which are the other kind of secondary
CR component.

The positron energy spectrum measured  recently in PAMELA, Fermi and AMS-02
experiments at kinetic energy $\epsilon_{k} > 10$~GeV turned out to be much 
flatter
than it is expected for positrons created in p-p collisions in ISM.
This stimulated many kind of assumptions that significant part of positrons
originate from new astrophysical sources such as pulsars or  
the annihilation of dark matter particles \citep[see][for a review]{serpico}.

At the same time SNRs are not only the most probable sources of primary CRs
with energies below $~10^{17}$~eV \citep[e.g.][]{bv07},
but also contribute significantly in the production of secondary CRs
due to reacceleration of already existing in ISM CRs and due to nuclear
collisions of primary CR particles with gas nuclei leading to the
creation of secondary particles which undergo subsequent acceleration by SNR shock.
The detailed study of these processes for the case of secondary nuclei \citep{bkpvz03}
based on the nonlinear kinetic theory of CR acceleration in SNRs \citep{bek96}
demonstrated that SNRs are expected to
contribute significantly to secondary CR spectra 
at kinetic energies $\epsilon_{k} > 10$~GeV/nucleon
making it significantly flatter.
Therefore it is natural to suggest that the observed flattering of positron
energy spectrum at $\epsilon_{k} > 10$~GeV 
can be partly or even predominantly due to SNR contribution.
Simple estimate \citep{blasi09} and the detailed study \citep{bk13}
have indeed confirmed that
the observed flat high energy positron spectrum is consistent with 
the expected SNR contribution.
(Note that both these studies were performed in the leaky-box framework which is 
a poor approximation for electrons and positrons. More detailed consideration, based on the diffusive
model for CR prpopagation in the ISM \citep{ahlers09, mertsch14},
confirmed such a conclusion.)
One can therefore  expect that SNRs also contribute significantly
to antiproton spectrum at high energies.

Here we calculate the energy spectrum of antiprotons
produced in SNRs to compare it with the existing data and  make the prediction at higher
energies $\epsilon_{k} > 100$~GeV where experimental data are not existed yet.
As a consistency check \citep[see also][for a similar point]{mertsch} we have also calculated
within the same model the energy spectra of secondary nuclei
and show that measured boron-to-carbon ratio 
is consistent with the considerable  SNR contribution.

\section{Production of secondary CRs in SNRs}
Acceleration of CRs in SNRs starts at some relatively low 
energy when some kind of suprathermal particles begin to cross the
SNR shock front. Any mechanism which supply suprathermal particles
into the shock acceleration is called injection.

Some small fraction of the
postshock thermal particle population are able to recross the shock
that means the beginning of their shock acceleration.
This is the most general and the most intense injection mechanism.
It occurs for all kind of ions and electrons existing
in the interstellar medium (ISM) and therefore it is
relevant for primary CRs only.
The corresponding injection rate is determined by the number of
particles involved into the acceleration from each medium volume
crossed the shock
and can be represented in the form \citep{bek96}:
\begin{equation}
N_\mathrm{inj}= \eta N_{g1} , ~
p_\mathrm{inj}=\lambda m c_{s2},  
\label{eq1}
\end{equation} 
where  $N_\mathrm{g}=\rho/m_\mathrm{p}$ is the gas number density, 
$c_\mathrm{s}$ is the sound speed,
the subscripts 1(2) refer to the point just ahead (behind) the shock.
Typical values of the dimensionless injection parameters
which provide CR production with required efficiency are
$\eta = 3\times 10^{-4}$ and $\lambda =4$.  Secondary CRs  like 
secondary nuclei
Li, Be, B or positrons and antiprotons by definition are not presented
in ISM and therefore they can not be produced due to such an injection.

Kinetic energy of all kind of Galactic cosmic ray (GCR) particles 
is considerably larger then the energy of gas particles injected
from the postshock thermal pool. Therefore
all GCRs which meet the expanding SNR shock are naturally involved
into the diffusive shock acceleration.
CR acceleration due to this  second relevant injection mechanism  is usually
called "reacceleration".
This term is used to distinguish the process of further increase of energy
of already energetic particles due to  interactions with SNR shocks
during their propagation in ISM. In this regard it is similar to the
stochastic acceleration (also called reacceleration) of GCRs due to
their interactions with background MHD-turbulence.
Since GCR energy spectra are relatively steep
and it have a peak at kinetic energy
$\epsilon_{k} =\epsilon_\mathrm{GCR}\sim 1$~GeV  
their injection can be represented in the form
\begin{equation}
N_\mathrm{inj}=N_\mathrm{GCR},~
 p_\mathrm{inj}= p_\mathrm{GCR},
\label{eq2}
\end{equation}
where $N_\mathrm{GCR}$ is the total number of GCR species per unit volume and 
$p_\mathrm{GCR}$
is their mean momentum, that corresponds to $\epsilon_\mathrm{GCR}$.
 
Primary  nuclei during their acceleration inside SNRs
produce  secondary nuclei in  nuclear collisions with the background
gas like GCRs do it in the Galactic disk. Essential fraction
of these already energetic particles has possibility to
be involved in further shock acceleration.
This is the third mechanism of secondary CR production inside SNRs.
For the first time it was studied to
describe the formation of the secondary CR nuclei spectra \citep{bkpvz03}. 

The production rate of secondary CR particles
due to the nuclear collisions of primary CRs with the gas nuclei within SNR
can be described
by the source term
\begin{equation}
q_{s}(r, p, t)=
4\pi cN_{g} \int_0^{\infty} 
dp'p'^2
\sigma_\mathrm{ps}(p, p') 
f_{p}(r, p',t),
\label{eq3}
\end{equation}
in the transport equation for the distribution function 
of secondary CRs $f_{s}(r, p',t)$. Here $\sigma_\mathrm{ps}(p, p')$ is
inelastic cross-section of secondary CRs production with momentum $p$ in
the collision of primary (parent) nuclei of momentum $p'$ with the gas nuclei,
$t$ is the time interval since the supernova explosion, $r$ is
the radial distance from the presupernova star. In the case of
secondary nuclei Li, Be, B the parent nuclei are heavier nuclei C, N, O,
whereas in the case antiprotons  all kind of accelerated
in SNRs nuclei (predominantly protons) play a role of parent nuclei.
Reacceleration and the acceleration of nuclei created in SNRs
in nuclear collisions are of the prime importance for the secondary CRs
even though relatively small part of primary CRs is also produced due to these 
processes.

The source term $q_{s}$ describes
the creation
of secondaries throughout the remnant, everywhere downstream
and upstream of SNR shock up to the distances $d\sim l_{p}(p')$
of the order of the diffusive length $l_{p}(p')$ of their parent primary CRs.
Essential part of these particles are naturally involving in
the acceleration at SNR shock.
It includes all the particles created upstream and the particles
created downstream at distances less then their diffusive length
$l_{s}(p)$ from the shock front. 
The number of these particles is increasing function of
their momentum because $l\propto \kappa(p)\propto p$
for the Bohm type diffusion coefficient $\kappa(p)$ which is realized 
during efficient CR acceleration in SNRs \citep[e.g.][]{ber08}. This 
makes the secondary particle spectra 
\begin{equation}
N_{s}(p,t)=16\pi^2 p^2\int_0^{\infty}dr r^2 f_{s}(r,p,t),
\label{eq4}
\end{equation}
produced in SNR,
harder compared with
the spectra of primaries $N_{p}(p,t)$.

The SNR efficiently accelerates  CRs
up to some maximal age $T_\mathrm{SN}$
when SNR release all previously accelerated CRs, primaries and secondaries,
with the spectra 
$N_{p}(\epsilon_{k})=v^{-1}N_{p}(p,T_\mathrm{SN})$ and
$N_{s}(\epsilon_{k})=v^{-1}N_{s}(p,T_\mathrm{SN})$ respectively,
into surrounding ISM.
Here $\epsilon_{k}$ and $v$ are the kinetic energy  
and speed of particle with momentum $p$.

The number of secondary CRs involved into the reacceleration  
at the SNR evolutionary epoch $t$
is proportional to the SNR volume $V(t)$, therefore $N_{s}(p,t)\propto V(t)$.
This is not so for primary CRs because the progressively increasing number
of injected CRs $N_\mathrm{inj}V$ is accompanied by the decrease of
their momentum $p_\mathrm{inj}\propto V_\mathrm{s}$ due to the shock deceleration. 
As a result
the number of primary relativistic CRs $N_{p}(p,t)$ remains 
nearly constant at late Sedov evolutionary phases \citep[e.g.][]{bek96}.
As a consequence the number of secondary CRs created in  nuclear collisions
$N_{s}(p,t)\propto tN_{p}(p,t)$ increases at late evolutionary phases proportionally
to SNR age $t$. Due to above factors secondary CRs are manly produced
on the late SNR evolutionary phases $t\sim T_\mathrm{SN}$.

These CRs released from SNRs together with
secondary CRs produced in ISM form 
the total secondary $n_{s}(\epsilon_{k})$ and primary 
$n_{p}(\epsilon_{k})$
CR populations. At sufficiently high energies the s/p ratio
of nuclear component
within simple leaky box model
is given by the expression \citep{bkpvz03}
\begin{equation}
\frac{n_{s}}{n_{p}}=\frac{n'_{s}}{n_{p}}+
\frac{N_{s}}{N_{p}},
\label{eq5}
\end{equation}
where 
$n'_{s}(\epsilon_{k})$ represents the spectrum of 
secondaries produced in nuclear collisions of
primary CRs within the Galactic disk.
Within the  leaky box model it
is approximately given by the expression \citep{bkpvz03}
${n'_{s}}/{n_{p}} = \sigma x/m_{p}$, where 
$x=\rho v\tau_\mathrm{esc}$ 
is the escape length which is the mean matter thickness traversed by GCRs in 
the course of their random walk in the Galaxy,
$\rho$ is the ISM gas density, 
$\tau_\mathrm{esc}(\epsilon_{k})$ 
is the CR escape time from the Galaxy, $m_{p}$ is the proton mass,
\begin{equation}
\sigma=\int_0^{\infty} 
d\epsilon_{k}\sigma_\mathrm{ps}(\epsilon_{k},\epsilon'_{k})
n_{p}(\epsilon'_{k})/
\int_0^{\infty} 
d\epsilon_{k}n_{p}(\epsilon'_{k}).
\label{eq7}
\end{equation}

Note that at sufficiently high energies 
the s/p ratio $n_\mathrm{s}/n_\mathrm{p}
\approx N_\mathrm{s}/N_\mathrm{p}$ is determined
by the s/p ratio  $N_\mathrm{s}/N_\mathrm{p}$ produced in the SNRs
independently on the propagation model which influence the ratio
${n'_{s}}/{n_{p}}$.

\section{Results and discussion}

We
have calculated the overall energy spectra of 
all relevant CR species accelerated in SNRs,
within  kinetic nonlinear model. 
The model is based on a fully time-dependent self-consistent solution of the CR
transport equation together with the gas dynamic equations in spherical
symmetry. It includes
the most relevant physical factors, essential for the evolution
and CR acceleration in a SNR and it is able to make quantitative predictions of the expected
properties of CRs produced in  SNRs and their nonthermal
radiation. The application of the theory to individual
SNRs  has demonstrated its capability of explaining the observed SNR
nonthermal emission properties \citep{ber08}. The theory
is able to explain 
major characteristics of the observed CR spectrum up to an energy of $\sim
10^{17}$~eV \citep[see][for details]{bek96, ber08}. Similar approach
was developed recently by other authors \citep{pzs10, kang10}.

We restrict ourselves by the most simple case of type Ia SN  in uniform ISM
with corresponding SN parameter values: 
explosion energy $E_\mathrm{SN}=10^{51}$~erg  and
ejecta mass$M_\mathrm{ej}=1.4  M_{\odot}$. 
We use typical values of the dimensionless parameters  $\eta=3\times10^{-4}$ and $\lambda=4$,
which describes the injection of gas particles into the shock acceleration \citep{vbk03}.
We consider the typical 
ISM with hydrogen
number density 
$N_{H}=1.5$~cm$^{-3}$, temperature $T_0=10^4$~K and magnetic field values
5~$\mu$G,
which roughly corresponds to the average ISM within the Galactic disk.
We adopt time-independent upstream magnetic field value $B_0$
and ignore magnetic field amplification effect
because  the secondaries are mainly produced at the
late evolutionary phases \citep{bkpvz03} when this effect becomes irrelevant.

We perform selfconsistent calculation up to the SNR age $T_\mathrm{SN}$
when SNR release all previously accelerated CRs into surrounding ISM.
We adopt the  value $T_\mathrm{SN}=10^5$~yr appropriate for considered range 
of ISM density \citep{bkpvz03}.

\begin{figure}
\plotone{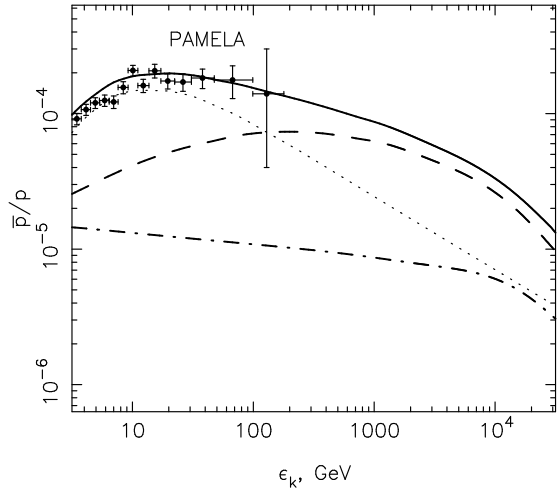}
\figcaption{\label{Fig1}
Calculated antiproton-to-proton ratio as a function of energy
together with PAMELA \citep{pamela10} data.
Dotted  and dashed  lines corresponds spectra of 
antiprotons created  in p-p collisions
in ISM \citep{donato01} and inside SNR respectively, dash-dotted 
line corresponds to the spectrum of antiprotons
reaccelerated in SNRs, solid line represent the sum of
contributions of all these processes. 
 }
\end{figure}

Calculated antiproton-to-proton ratio 
$\bar p/p=n_{\bar p}(\epsilon_{k})/ n_{p}(\epsilon_{k})$ as a function of 
energy
together with PAMELA data are shown in Fig.~\ref{Fig1}.
We use the crossection 
$\sigma(\epsilon, \epsilon') = 1.6\sigma_{pp}(\epsilon, \epsilon')$, 
parametrization of \citet{shibata} for crossection of antiproton production
in p-p collisions $\sigma_{pp}(\epsilon, \epsilon')$ and correction factor 1.6 which
describes the contribution of heavier nuclei.
Since the observed antiproton spectrum has a peak at $\epsilon_{k}\approx 2$~GeV
we use $N_\mathrm{inj}=10^{-14}$~cm$^{-3}$ and the value of $p_\mathrm{inj}$
which corresponds to the kinetic energy $\epsilon_{k}= 2$~GeV.
For the spectrum of antiprotons produced in ISM we use the results 
of calculations performed in \citet{donato01}. It is very close to what was used
by \citet{BlasiSer09} in their similar consideration.
It is seen that  antiprotons at
energies $\epsilon_{k}<10$~GeV are produced in SNRs  
equally effectively by both mechanisms 
whereas at $\epsilon_{k}>10$~GeV
the creation of antiprotons 
in p-p collisions
and their subsequent acceleration becomes dominant.
In total the antiproton production in SNRs makes
the  energy dependence of $\bar p/p$
considerably more flatter so that at $\epsilon_{k}\sim 10^3$~GeV
the ratio becomes  larger by a factor of about five.
Within the energy range 30~GeV~$<\epsilon_{k}< 3000$~GeV
the energy dependence of the ratio $\bar p/p\propto \epsilon_{k}^{-0.25}$ 
is expected to be very flat.

PAMELA data, which well agrees with our calculation,
within the energy range 10~GeV~$<\epsilon_{k}< 100$~GeV
provide the evidence that the actual  ratio $\bar p/p$ is indeed
more flatter than it is expected if antiprotons are created in ISM only.

%
\begin{figure}
\plotone{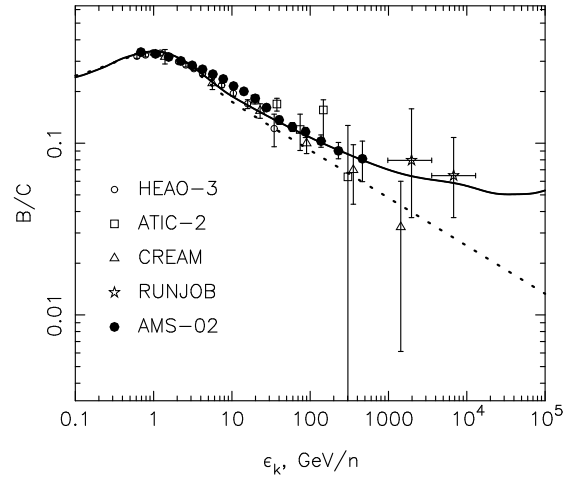}
\figcaption{\label{Fig2}
Calculated boron-to-carbon  ratio as a function of 
kinetic energy per nucleus
together with the results of
HEAO-3 \citep{heao3}, ATIC-2 \citep{atic2}, CREAM \citep{cream},
RUNJOB\citep{runjob} and AMS-02 \citep{ams02} experiments.
Dotted line corresponds to spectrum of boron created  in nuclear collisions
in ISM \citep{bkpvz03}, solid line represents the boron spectrum which
includes the contribution of SNRs. 
 }
\end{figure}

The production of antiprotons in SNRs estimated by 
\citet{BlasiSer09} is considerably larger (by a factor of four at  
$\epsilon_{k}=10^3$ ~GeV)
compared with our calculation even though these authors have neglected
the reacceleration process at all. It is due to a number of 
simplifications made by these authors.
For example  in actual situation the
overlap between the radial profile of protons 
$f(r,p)$ with the gas density profile $\rho(r)$ which has a
peak value $\rho\approx 4 \rho_\mathrm{ISM}$ at the shock $(r=R_\mathrm{s})$, progressively
decreases with increase of energy at high energies $\epsilon_{k}>100$ ~GeV,
because the radial profile of protons becomes progressively
broader. Here $\rho_\mathrm{ISM}$ is the gas density of ISM.
This leads to the decrease of the effective gas density  from the
value $\rho(r=R_\mathrm{s})\approx 4 \rho_\mathrm{ISM}$ to $\rho(r>R_\mathrm{s})\approx  
\rho_\mathrm{ISM}$  and to $\rho(r<R_\mathrm{s})<  
\rho_\mathrm{ISM}$
that follows by the decrease of the antiproton production.
This factor was neglected by \citet{BlasiSer09},
that is one of the reason which lead them to overestimation of antiproton production.

In order to check the consistency of other types of secondary CR
production we have calculated within the same model
the boron-to-carbon (B/C)  ratio
and compare it in Fig.~\ref{Fig2}  with the 
existing experimental data.

The boron nuclei represent the example of secondary nuclei.
To calculate boron  spectrum we use the overall number density
of boron nuclei in ISM $N_\mathrm{inj}=7.9\times
10^{-14}$~cm$^{-3}$ 
injected at a kinetic energy $\epsilon_\mathrm{inj}=0.6$~GeV/n which
corresponds to the mean GCR energy for these element. 
Compared with the previous study \citep{bkpvz03}
except C, N and O nuclei as parent species following \citet{mertsch} and \citet{td12}
we included
also  heavier primaries up to Si,
which contribute about 10\% into the boron production.

Due to boron production in SNRs the expected  B/C ratio 
undergoes considerable flattering which starts  at energy $\epsilon_{k}\approx 
100$~GeV/nucleon.
As one can  see in Fig.~\ref{Fig2} this is  consistent with the 
measurements recently
performed in balloon \citep{runjob} and AMS-02  space \citep{ams02} experiments
even though for more strict conclusion one needs the measurements with higher statistics
at energies above 1~TeV/n.

Calculated SNR contribution into the secondary CR spectra represents the component
which is unavoidably expected if SNRs are the main source of GCRs.
Comparison with the existing data leads to a conclusion 
that the observed high energy excess of  secondary nuclei
can be  
produced in Galactic SNRs. This enable to expect  similar
excess in the antiproton energy spectrum. The data expected very soon from AMS-02
experiment will make it clear whether the actual ratio  $\bar p/p$
is indeed not less flat at energies $\epsilon_{k}> 10$~GeV then we predict.

\begin{acknowledgments}
This work is supported by the Russian Foundation for Basic Research (grants
13-02-00943 and 13-02-12036) and by the Council of the
President of the Russian Federation for Support of Young Scientists and Leading
Scientific Schools (project No. NSh-3269.2014.2).
\end{acknowledgments}

\end{document}